\documentclass[a4paper]{article}
\usepackage[english]{babel}
\usepackage[OT1]{fontenc}
\usepackage[utf8]{inputenc}
\usepackage{amsmath} 
\usepackage{amsfonts}
\usepackage{amssymb}
\usepackage{amsthm}
\usepackage{latexsym}
\usepackage{cancel}

\usepackage{amscd}
\usepackage{graphicx}
\usepackage{fancyhdr}
\usepackage{pbox}

\providecommand{\pacs}[1]{PACS numbers : #1}
\providecommand{\keywords}[1]{Keywords : #1}

\usepackage[margin=1.2in,footskip=0.5in]{geometry}

\begin{document}
\title{Exact form of the generalized Lorentz force in Fock's nonlinear relativity}
\author{N. Takka\hspace{0.1cm}\footnote{E-mail: takka.naimi@gmail.com} \\  Laboratoire de Physique Théorique, Faculté des Sciences Exactes,\\
Université de Bejaia, 06000 Bejaia, Algeria}
\date{\today}

\maketitle

\begin{abstract}
This work completes a serie of two papers devoted to the extension of the fundamental laws of electrodynamics in the context of Fock's nonlinear relativity (FNLR). Indeed, after having established in the previous study the exact generalizations of both Maxwell's equations and Dirac magnetic monopole, we present here the remaining exact Lorentz force. As in $\kappa$-Minkowski spacetime, two different nature contributions appear in the corresponding equation of motion where the new effect is interpreted as the gravitational-type Lorentz force. This common point separately induced by the radius of the universe in our case or Planck energy in other works, reinforces once more the analogy between electromagnetism and gravity in two different scientific approaches. As a relative difference, it is very important to highlight that more homogeneity characterizes our results where each effect is exclusively generated by mass or charge but not both at the same time. Even more, the new effect emerges as the result of the triple effect of $R$-deformation, mass and the square of velocities but completely independent of electromagnetic field. 
\end{abstract}

        

\newpage

 
 

\section{Introduction}

Among many approaches to present the laws of electrodynamics, Feynman’s post-humous paper, published by Freeman J. Dyson in $1990$, constitutes a very elegant mathematical way for such derivation\cite{Dyson-1}. The initial motivation behind that curiosity was to go as far as possible beyond the framework of conventional physics in order to develop a quantization procedure without resorting to a Lagrangian or Hamiltonian. In its original version, this approach is based on the commutation relations between position and velocity for a nonrelativistic particle and Newton’s law of motion valid in three-dimensional Euclidean space. To reconcile the aforementioned proof with some conflicts between Lorentz covariance and the Euclidean assumptions
raised in Refs. \cite{Dombey-Brehme-Anderson-Farquhar-Hughes}-\cite{Vaidya-Farina-1}, Tanimura formulated the corresponding special and general relativistic versions \cite{Tanimura-1}. In this latter, it has also shown that the only possible fields
that can consistently act on a quantum particle are scalar, gauge and gravitational fields. Some years later, the use of Hodge duality made possible the derivation of the two groups of Maxwells equations with a magnetic monopole in flat and
curved spaces \cite{Berard-Grandati-Mohrbach-1}. In many other works, Bérard et al. also established a connection
between Lorentz algebra symmetry and Dirac’s magnetic monopole, e.g. Refs. \cite{Berard-Grandati-Mohrbach-2} and \cite{Berard-Mohrbach-1}. Thinking in terms of analogy, the equivalence between Feynman’s proof and minimal coupling has been discussed in commutative space–time \cite{Montesinos-Perez-Lorenzana-1}. Moving forward to noncommutative space–time, it is turned out that the extended first approximation of the laws of electrodynamics in $\kappa$-Minkowski space–time depends on the particle mass \cite{Harikumar-1} and \cite{Harikumar-Juric-Meljanac-1}. Here of course, $\kappa$ represents Planck energy divided by the speed of light in a vacuum. In Ref. \cite{Harikumar-Juric-Meljanac-1} this is done by power series expansion of noncommutative coordinates, momenta and some derived functions, in terms of commutative coordinates, momenta and the deformation parameter $\kappa$. Finally, we have suggested in our turn a new version of Feynman’s proof within which the electromagnetism can be studied formally \cite{Takka-Bouda-Foughali-1}. With such reasoning, we have then found the first-order approximation of Maxwell’s equations, Lorentz force and Dirac’s magnetic monopole in FNLR \cite{Takka-Bouda-Foughali-1}. A year later and except for the equation of motion, we have succeeded to go further than our first paper by establishing the corresponding exact formulations in the second one \cite{Takka-Bouda-2}. Calculating the flux of a magnetostatic field generated by the hypothetical magnetic charge across a closed surface, it is turned out that the universe could locally contain the magnetic charge but in its totality it is still neutral. Now, to talk briefly about FNLR, let us take as origin \cite{Fock-1} where the general form of the coordinates transformation between two inertial frames, based only on the first principle of relativity, is given by the formula

\begin{equation}\label{equation-1}
t^{\prime }=\frac{\gamma\big(t-ux/c^2\big)}{\alpha_R},\qquad x^{\prime }=\frac{
\gamma(x-ut)}{\alpha_R},\qquad y^{\prime }=\frac{y}{\alpha_R},\qquad z^{\prime}=\frac{z}{\alpha_R},
\end{equation}

\noindent here obviously $\gamma =\big(1-u^{2}/c^{2}\big)^{-1/2}$ and ${\alpha _{R}}=1+\left[(\gamma -1)ct-\gamma xu/c\right]/R$ where $R$ designates the radius of the universe. To obtain a coherent theory, this latter has undergone several steps. In fact, unlike deformed (or doubly) special relativity (DSR), whose construction is based on three postulates  by imposing a maximal energy to special relativity \cite{Amelino-1}-\cite{Magueijo-Smolin-2}, FNLR takes the diametrically opposed direction by studying the implications of the non-constancy of the speed of light in vacuum.  Subsequently, this trend is greatly reinforced by the desire to resolve some problems and paradoxes in cosmology \cite{Albrecht-Magueijo-1} and in quantum gravity \cite{Magueijo-Smolin-1}. A few years later, a further step in that perspective has been realized by inspiring from one of the last versions of DSR \cite{Ghosh-Pal-1}. In other words, the usual Fock-Lorentz transformation (\ref{equation-1}) has been reproduced and a new momentum transformation suggested  as a result of the definition of the appropriate deformed Poisson brackets \cite{Bouda-Foughali-1}. The main novelty of such rewriting is that the contraction $x_{\mu}p^{\mu}$ becomes an invariant which made possible the coherent description of free particles in this context. Focusing on the generalizations of both Klein-Gordon and Dirac equations in \cite{Foughali-Bouda-1} and \cite{Foughali-Bouda-2} respectively, the correspondence between $R$-Minkowski and de Sitter spacetime has been highlighted.\\

This paper is organized as follows. In section $2$, we  briefly recall some needed results found in \cite{Takka-Bouda-2} in order to find the exact Lorentz force in FNLR. To this end, we have used one of the common points between our version of Feynman's proof and that developed by Tanimura \cite{Tanimura-1}, namely, the relativistic Newton's law. As consequence, a partial dependance on particle mass emerges naturally in the corresponding equation of motion. Section $3$ is devoted to conclusion where a special case is discussed.

 
 

\section{Exact $R$-Lorentz force}
 To begin, let us recall that the quantized $R$-deformed phase space algebra is described by the following three commutators \cite{Foughali-Bouda-1} 

\begin{align}
\label{equation-2}
\left[x^\mu,x^\nu\right] & = 0, \\
\label{equation-3}
\left[x^\mu,p^\nu\right] & = - i \hbar \eta^{\mu\nu} + \frac{i \hbar}{R}\eta^{0\nu}x^\mu, \\
\label{equation-4}
\left[p^\mu,p^\nu\right] & = -\frac{i \hbar}{R}\left(p^\mu\eta^{0\nu}-p^\nu\eta^{\mu0}\right),
\end{align}

\noindent where $\eta^{\mu\nu} = (+1,-1,-1,-1)$ and $\mu ,\nu =0,1,2,3$. To derive the generalized fundamental laws of electrodynamics in the context of FNLR, we have developed in \cite{Takka-Bouda-Foughali-1} a new relativistic version of Feynman's proof. This latter is based on the knowledge of phase space algebra and the explicit form of the four-momentum valid in the absence of electromagnetic field. To this end, we were inspired by the equivalence between the minimal coupling prescription and Feynman’s approach highlighted in the first paper \cite{Dyson-1} and confirmed in further works, e.g. \cite{Montesinos-Perez-Lorenzana-1} and \cite{Harikumar-Juric-Meljanac-1}. On a quest for new generalizations even more satisfying, we set up a symmetrization mechanism giving the following exact formulations of  momentum and commutation relations respectively \cite{Takka-Bouda-2}

\begin{equation}\label{equation-5}
p^{\mu}=m\Big(1-\frac{x^{0}}{R}\Big)^{-3}\left\{\dot{x}^{\mu}+\frac{1}{R}\big(x^{\mu}\dot{x}^{0}-x^{0}\dot{x}^{\mu}\big)\right\}+\frac{3i\hbar}{2R}\eta^{\mu 0},
\end{equation}

\begin{equation}\label{equation-6}
[x^{\mu}, \dot{x}^{\nu}]=\frac{i\hbar}{m}\Big(1-\frac{x^{0}}{R}\Big)^{2}\left\{-\eta^{\mu\nu}+\frac{1}{R}\big(\eta^{0\nu}x^{\mu}+\eta^{\mu 0}x^{\nu}\big)-\frac{x^{\mu}x^{\nu}}{R^{2}}\right\},
\end{equation}

\begin{equation}\label{equation-7}
[\dot{x}^{\mu}, \dot{x}^{\nu}]=-\frac{i\hbar q}{m^{2}}F^{\mu\nu}(x)-\frac{i\hbar}{mR}\Big(1-\frac{x^{0}}{R}\Big)^{2}\Big\{\big(\eta^{\mu 0}\dot{x}^{\nu}-\eta^{0\nu}\dot{x}^{\mu}\big)
-\frac{1}{R}\big(x^{\mu}\dot{x}^{\nu}-x^{\nu}\dot{x}^{\mu}\big)\Big\},
\end{equation}

\noindent obviously here $F^{\mu\nu}(x)$ describes the generalized electromagnetic tensor. The unique dependence of this latter on position has been deduced from the following result $\big[x^{\lambda}, F^{\mu\nu}\big]=0$. Now, by differentiating the commutator involving one coordinate and one velocity with respect to the time parameter $\tau$, the equation of motion describing a moving charged particle under the influence of electromagnetic field is given by \cite{Tanimura-1}

\begin{equation}\label{equation-8}
\left[x^{\mu}, F^{\nu}\right]=m\left[x^{\mu}, \ddot{x}^{\nu}\right]=m\frac{d}{d\tau}\left[x^{\mu}, \dot{x}^{\nu}\right]-m\left[\dot{x}^{\mu}, \dot{x}^{\nu}\right].
\end{equation}

\noindent Above, the relativistic Newton's law is used as an assumption allowing the recovery of the usual results of special relativity. Given that the derivation of $F^{\nu}$ is conditioned by the knowledge of the two intervening commutators,  to determine $\frac{d}{d\tau}\left[x^{\mu}, \dot{x}^{\nu}\right]$, we propose below a mechanism giving such calculation. First of all, we begin by taking the time derivative of (\ref{equation-6}), to get

\begin{eqnarray}\label{equation-9}
m\frac{d}{d\tau}\left[x^{\mu}, \dot{x}^{\nu}\right]&=&\frac{2i\hbar}{R}\eta^{\mu\nu}\Big<\left(1-\frac{x^{0}}{R}\right)\dot{x}^{0}\Big>-\frac{2i\hbar}{R^{2}}\left\{\eta^{0 \nu}\Big<\left(1-\frac{x^{0}}{R}\right)x^{\mu}\dot{x}^{0}\Big>\right.\nonumber\\
                                                   &+& \left.\eta^{\mu 0}\Big<\left(1-\frac{x^{0}}{R}\right)x^{\nu}\dot{x}^{0}\Big>\right\} + \frac{2i\hbar}{R^{3}}\Big<\left(1-\frac{x^{0}}{R}\right)x^{\mu}x^{\nu}\dot{x}^{0}\Big>\nonumber\\
                                                   &+&\frac{i\hbar}{R}\left\{\eta^{\mu 0}\Big<\left(1-\frac{x^{0}}{R}\right)^{2}\dot{x}^{\nu}\Big>+\eta^{0\nu}\Big<\left(1-\frac{x^{0}}{R}\right)^{2}\dot{x}^{\mu}\Big>\right\}\nonumber\\
                                                   &-&\frac{i\hbar}{R^{2}}\left\{\Big<\left(1-\frac{x^{0}}{R}\right)^{2}x^{\mu}\dot{x}^{\nu}\Big> + \Big<\left(1-\frac{x^{0}}{R}\right)^{2}x^{\nu}\dot{x}^{\mu}\Big>\right\},     
\end{eqnarray}

\noindent  where $<.>$ refers to the symmetrization operator \cite{Tanimura-1}. To explicit each of the above terms, we proceed one by one. First, starting from the definition of the commutator 

\begin{equation}\label{equation-commutator}
\dot{x}^{\mu}x^{0}=x^{0}\dot{x}^{\mu}-[x^{0}, \dot{x}^{\mu}],
\end{equation}

\noindent where

\begin{equation}\label{equation-10}
[x^{\mu}, [x^{\nu}, \dot{x}^{\lambda}]]=0
\end{equation}

\noindent and

\begin{equation}\label{equation-interchangeability}
[x^{\mu}, \dot{x}^{\nu}]=[x^{\nu}, \dot{x}^{\mu}],
\end{equation}

\noindent we get the following generalization \cite{Takka-Bouda-2}

\begin{equation}\label{equation-11}
\big<f(x^{0})\dot{x}^{\mu}\big>=f(x^{0})\dot{x}^{\mu}-\frac{1}{2}\frac{d f(x^{0})}{dx^{0}}[x^{0}, \dot{x}^{\mu}],
\end{equation}

\noindent which, by means of (\ref{equation-6}), gives

\begin{equation}\label{equation-12}
\frac{2i\hbar}{R}\eta^{\mu\nu}\Big<\left(1-\frac{x^{0}}{R}\right)\dot{x}^{0}\Big> = \frac{2i\hbar}{R}\eta^{\mu\nu}\left(1-\frac{x^{0}}{R}\right)\dot{x}^{0} + \frac{\hbar^{2}}{mR^{2}}\eta^{\mu\nu}\left(1-\frac{x^{0}}{R}\right)^{4}.
\end{equation}  

\noindent To calculate the other two terms appearing in the same order in (\ref{equation-9}), let us recall that for three operators $A$, $B$, and $C$ satisfying the following commutation relations 

\begin{equation}\label{equation-13}
 [A,B]=0\hspace{0.2cm}\text{and}\hspace{0.2cm}[A,[B,C]]=0,
\end{equation}

\noindent we have \cite{Takka-Bouda-2}

\begin{equation}\label{equation-14}
\big<ABC\big>=\frac{1}{2}A\big<BC\big>+\frac{1}{2}\big<BC\big>A
\end{equation}

\noindent and then the second term can be written as

\begin{equation}\label{equation-15}
\Big<\left(1-\frac{x^{0}}{R}\right)x^{\mu}\dot{x}^{0}\Big>= \frac{1}{2}\left(1-\frac{x^{0}}{R}\right)\left<x^{\mu}\dot{x}^{0}\right> + \frac{1}{2}\left<x^{\mu}\dot{x}^{0}\right> \left(1-\frac{x^{0}}{R}\right).
\end{equation}

\noindent Using Eq. (\ref{equation-6}), we can make sure that

\begin{equation}\label{equation-16}
\left<x^{\mu}\dot{x}^{0}\right>
                               = x^{\mu}\dot{x}^{0} + \frac{1}{2}\left[\dot{x}^{0}, x^{\mu}\right] 
                               = x^{\mu}\dot{x}^{0} + \frac{i\hbar}{2m}\left(1-\frac{x^{0}}{R}\right)^{3}\left(\eta^{\mu 0}-\frac{x^{\mu}}{R}\right)
\end{equation}

\noindent and similarly

\begin{equation}\label{equation-17}
\dot{x}^{0} \left(1-\frac{x^{0}}{R}\right)=\left(1-\frac{x^{0}}{R}\right)\dot{x}^{0}-\frac{i\hbar}{mR} \left(1-\frac{x^{0}}{R}\right)^{4},
\end{equation}

\noindent which, after substitution into (\ref{equation-15}), yields

\begin{eqnarray}\label{equation-18}
-\frac{2i\hbar}{R^{2}}\eta^{0 \nu}\Big<\left(1-\frac{x^{0}}{R}\right)x^{\mu}\dot{x}^{0}\Big>= &-& \frac{2i\hbar}{R^{2}}\eta^{0 \nu}\left(1-\frac{x^{0}}{R}\right) x^{\mu}\dot{x}^{0} -\frac{2\hbar^{2}}{mR^{3}} \eta^{0 \nu}\left(1-\frac{x^{0}}{R}\right)^{4}x^{\mu}\nonumber\\
&+& \frac{\hbar^{2}}{mR^{2}}\left(1-\frac{x^{0}}{R}\right)^{4} \eta^{\mu 0} \eta^{0 \nu}.
\end{eqnarray}

\noindent After interchanging $\mu$ with $\nu$ in the last result, the third term is given by

\begin{eqnarray}\label{equation-19}
-\frac{2i\hbar}{R^{2}}\eta^{\mu 0}\Big<\left(1-\frac{x^{0}}{R}\right)x^{\nu}\dot{x}^{0}\Big>= &-& \frac{2i\hbar}{R^{2}} \eta^{\mu 0}\left(1-\frac{x^{0}}{R}\right) x^{\nu}\dot{x}^{0} -\frac{2\hbar^{2}}{mR^{3}} \eta^{\mu 0 }\left(1-\frac{x^{0}}{R}\right)^{4}x^{\nu}\nonumber\\
&+& \frac{\hbar^{2}}{mR^{2}}\left(1-\frac{x^{0}}{R}\right)^{4} \eta^{\mu 0} \eta^{0 \nu}. 
\end{eqnarray}

\noindent To continue, we  generalize in its turn Eq. (\ref{equation-14}). Indeed, for four operators $A$, $B$ $C$ and $D$ satisfying the following commutation relations 

\begin{equation}\label{equation-20}
 [A,B]=[B,C]=0\hspace{0.2cm}\text{and}\hspace{0.2cm}[A,[B,D]]=[A,[C,D]]=0,
\end{equation}

\noindent  we get

\begin{equation}\label{equation-21}
\big<ABCD\big>=\frac{1}{4}\left[ AB\big<CD\big>+B\big<CD\big>A+A\big<CD\big>B+\big<CD\big>AB\right].
\end{equation}

\noindent Identifying the last relation with the four term listed into (\ref{equation-9}), we obtain

\begin{eqnarray}\label{equation-22}
\Big<\left(1-\frac{x^{0}}{R}\right)x^{\mu}x^{\nu}\dot{x}^{0}\Big> &=& \frac{1}{4}\left[\left(1-\frac{x^{0}}{R}\right)x^{\mu}\left<x^{\nu}\dot{x}^{0}\right> + x^{\mu}\left<x^{\nu}\dot{x}^{0}\right>\left(1-\frac{x^{0}}{R}\right)\right.\nonumber\\
 && \hspace*{0.2cm}\left. +\left(1-\frac{x^{0}}{R}\right)\left<x^{\nu}\dot{x}^{0}\right>x^{\mu}+ \left<x^{\nu}\dot{x}^{0}\right>\left(1-\frac{x^{0}}{R}\right)x^{\mu}\right].
\end{eqnarray}

\noindent With the same reasoning, we can check that the use of Eqs. (\ref{equation-6}), (\ref{equation-commutator}), (\ref{equation-16}) and (\ref{equation-17}) transforms the last relation as follows

\begin{eqnarray}\label{equation-23}
\frac{2i\hbar}{R^{3}}\Big<\left(1-\frac{x^{0}}{R}\right)x^{\mu}x^{\nu}\dot{x}^{0}\Big>&=&\frac{2i\hbar}{R^{3}} \left(1-\frac{x^{0}}{R}\right)x^{\mu}x^{\nu}\dot{x}^{0} - \frac{\hbar^{2}}{mR^{3}}\left(1-\frac{x^{0}}{R}\right)^{4} \left[\left(\eta^{\mu 0}x^{\nu} + \eta^{0\nu}x^{\mu}\right)\right.\nonumber\\
&-& \left. \frac{3}{2}x^{\mu}x^{\nu}\right].
\end{eqnarray}

\noindent After that, taking into account Eqs. (\ref{equation-6}) and (\ref{equation-11}), we can make sure that the fifth and sixth  terms are respectively  written as

\begin{equation}\label{equation-24}
\frac{i\hbar}{R}\eta^{\mu 0}\Big<\left(1-\frac{x^{0}}{R}\right)^{2}\dot{x}^{\nu}\Big> = \frac{i\hbar}{R}\eta^{\mu 0}\left(1-\frac{x^{0}}{R}\right)^{2}\dot{x}^{\nu}
 + \frac{\hbar^{2}}{mR^{2}}\left(1-\frac{x^{0}}{R}\right)^{4} \eta^{\mu 0}\left(\eta^{0\nu}-\frac{x^{\nu}}{R}\right)
\end{equation}

\noindent and

\begin{equation}\label{equation-25}
\frac{i\hbar}{R}\eta^{0 \nu}\Big<\left(1-\frac{x^{0}}{R}\right)^{2}\dot{x}^{\mu}\Big> = \frac{i\hbar}{R}\eta^{ 0 \nu}\left(1-\frac{x^{0}}{R}\right)^{2}\dot{x}^{\mu}
 + \frac{\hbar^{2}}{mR^{2}}\left(1-\frac{x^{0}}{R}\right)^{4} \eta^{0 \nu}\left(\eta^{ \mu 0}-\frac{x^{\mu}}{R}\right).
\end{equation}

\noindent To find the seventh one, making use of Eq. (\ref{equation-14}) to get

\begin{equation}\label{equation-26}
\Big<\left(1-\frac{x^{0}}{R}\right)^{2}x^{\mu}\dot{x}^{\nu}\Big>= \frac{1}{2}\left(1-\frac{x^{0}}{R}\right)^{2}\left<x^{\mu}\dot{x}^{\nu}\right> + \frac{1}{2}\left<x^{\mu}\dot{x}^{\nu}\right> \left(1-\frac{x^{0}}{R}\right)^{2}.
\end{equation}

\noindent As for (\ref{equation-16}), we can check that

\begin{equation}\label{equation-27}
\left<x^{\mu}\dot{x}^{\nu}\right> = x^{\mu}\dot{x}^{\nu}+\frac{i\hbar}{2m}\Big(1-\frac{x^{0}}{R}\Big)^{2}\left\{\eta^{\mu\nu}-\frac{1}{R}\big(\eta^{0\nu}x^{\mu}+\eta^{\mu 0}x^{\nu}\big)+\frac{x^{\mu}x^{\nu}}{R^{2}}\right\}. 
\end{equation}

\noindent Exploiting Eqs. (\ref{equation-commutator}) and (\ref{equation-10}), we can prove by mathematical induction that for any regular function $f$ of $x^{0}$, we have

\begin{equation}\label{equation-29}
\dot{x}^{\nu}f(x^{0})=f(x^{0})\dot{x}^{\nu}-\frac{d f(x^{0})}{dx^{0}}[x^{0}, \dot{x}^{\nu}]
\end{equation}

\noindent and then

\begin{equation}\label{equation-30}
\dot{x}^{\nu}\left(1-\frac{x^{0}}{R}\right)^{2}=\left(1-\frac{x^{0}}{R}\right)^{2}\dot{x}^{\nu}-\frac{2i\hbar}{mR}\left(1-\frac{x^{0}}{R}\right)^{4}\left(\eta^{0 \nu}-\frac{x^{\nu}}{R}\right).
\end{equation}

\noindent  By means of the last result, the substitution of Eq. (\ref{equation-27}) into (\ref{equation-26}) yields

\begin{eqnarray}\label{equation-31}
-\frac{i\hbar}{R^{2}}\Big<\left(1-\frac{x^{0}}{R}\right)^{2}x^{\mu}\dot{x}^{\nu}\Big>=&-&\frac{i\hbar}{R^{2}}\left(1-\frac{x^{0}}{R}\right)^{2}x^{\mu}\dot{x}^{\nu}-\frac{\hbar^{2}}{mR^{3}}\left(1-\frac{x^{0}}{R}\right)^{4}\left(\eta^{0\nu}-\frac{x^{\nu}}{R}\right)x^{\mu}\nonumber\\
&+&\frac{i\hbar}{2R^{2}}\left(1-\frac{x^{0}}{R}\right)^{2}[x^{\mu}, \dot{x}^{\nu}],
\end{eqnarray}

\noindent which, after interchanging $\mu$ with $\nu$, gives the following last term

\begin{eqnarray}\label{equation-32}
-\frac{i\hbar}{R^{2}}\Big<\left(1-\frac{x^{0}}{R}\right)^{2}x^{\nu}\dot{x}^{\mu}\Big>=&-&\frac{i\hbar}{R^{2}}\left(1-\frac{x^{0}}{R}\right)^{2}x^{\nu}\dot{x}^{\mu}-\frac{\hbar^{2}}{mR^{3}}\left(1-\frac{x^{0}}{R}\right)^{4} \left(\eta^{\mu 0}-\frac{x^{\mu}}{R}\right)x^{\nu}\nonumber\\
&+&\frac{i\hbar}{2R^{2}}\left(1-\frac{x^{0}}{R}\right)^{2}[x^{\mu}, \dot{x}^{\nu}].
\end{eqnarray}

\noindent Finally, by substituting all intervening terms expressed by (\ref{equation-12}), (\ref{equation-18}), (\ref{equation-19}), (\ref{equation-23}), (\ref{equation-24}), (\ref{equation-25}), (\ref{equation-31}) and (\ref{equation-32})  into  (\ref{equation-9}), we get after simplification

\begin{eqnarray}\label{equation-33}
m\frac{d}{d\tau}\left[x^{\mu}, \dot{x}^{\nu}\right]&=&-\frac{2m}{R} \left(1-\frac{x^{0}}{R}\right)^{-1} [x^{\mu}, \dot{x}^{\nu}]\dot{x}^{0} + \frac{i\hbar}{R} \left(1-\frac{x^{0}}{R}\right)^{2}\left\{\left( \eta^{0\nu}\dot{x}^{\mu}+ \eta^{\mu 0} \dot{x}^{\nu}\right)\right.\nonumber\\
&&\left. -\frac{1}{R}\left( x^{\nu}\dot{x}^{\mu}+ x^{\mu} \dot{x}^{\nu}\right)+ \frac{1}{R}[x^{\mu}, \dot{x}^{\nu}]\right\} + \frac{\hbar^{2}}{mR^{2}}\left(1-\frac{x^{0}}{R}\right)^{4} \left\{\left( \eta^{\mu\nu} + 4\eta^{\mu 0} \eta^{0\nu}\right)\right.\nonumber\\
&&\left. -\frac{5}{R}\left( \eta^{0\nu}x^{\mu}+ \eta^{\mu 0} x^{\nu}\right) + \frac{5}{R^{2}}x^{\mu}x^{\nu} \right\}.
\end{eqnarray}

\noindent Taking into account (\ref{equation-7}) and (\ref{equation-33}), Eq.(\ref{equation-8}) gives

\begin{eqnarray}\label{equation-34}
\left[x^{\mu}, F^{\nu}\right]&=& \frac{i\hbar q}{m}F^{\mu\nu}(x)-\frac{2m}{R} \left(1-\frac{x^{0}}{R}\right)^{-1} \left\{[x^{\mu}, \dot{x}^{\nu}]\dot{x}^{0} + [x^{\mu}, \dot{x}^{0}]\dot{x}^{\nu}\right\}\nonumber\\
&+& \frac{i\hbar}{R^{2}} \left(1-\frac{x^{0}}{R}\right)^{2}[x^{\mu}, \dot{x}^{\nu}] + \frac{\hbar^{2}}{mR^{2}}\left(1-\frac{x^{0}}{R}\right)^{4} \left\{\left( \eta^{\mu\nu} + 4\eta^{\mu 0} \eta^{0\nu}\right)\right.\nonumber\\
&-&\left. \frac{5}{R}\left( \eta^{0\nu}x^{\mu}+ \eta^{\mu 0} x^{\nu}\right) + \frac{5}{R^{2}}x^{\mu}x^{\nu} \right\}.
\end{eqnarray}

\noindent Observing the last result where $[x^{\mu}, \dot{x}^{\nu}]=[x^{\mu}, \dot{x}^{\nu}](x)$, we remark that $\left[x^{\mu}, F^{\nu}\right]$ contains two kind of terms. One depends only on position and another is proportional to the velocity. As consequence,  Eqs. (\ref{equation-2}) and (\ref{equation-10}) indicate that the resulting force can be written as

\begin{equation}\label{equation-35}
F^{\nu}=q\big<\tilde{F}^{\nu\alpha}(x)\dot{x}_{\alpha}\big> + m\big< \Gamma^{\nu\alpha\beta}(x) \dot{x}_{\alpha} \dot{x}_{\beta}\big> + G^{\nu}(x)
\end{equation}

\noindent and then

\begin{equation}\label{equation-36}
\left[x^{\mu}, F^{\nu}\right]=q\tilde{F}^{\nu\alpha}(x)[x^{\mu}, \dot{x}_{\alpha}] + m\big< \Gamma^{\nu\alpha\beta}(x) \left\{ \left[x^{\mu}, \dot{x}_{\alpha}\right]\dot{x}_{\beta} + \dot{x}_{\alpha}\left[x^{\mu}, \dot{x}_{\beta}\right] \right\}\big>,
\end{equation}

\noindent which, after identification with Eq. (\ref{equation-34}), yields

\begin{equation}\label{equation-37}
\Gamma^{\nu\alpha\beta}(x)=-\frac{2}{R}\left(1-\frac{x^{0}}{R}\right)^{-1}\eta^{0\alpha}\eta^{\nu\beta}.
\end{equation}

\noindent Above $\Gamma^{\nu\alpha\beta}(x)$ represents the electromagnetic-type Christoffel symbols and $G^{\nu}(x)$ is just an integration function depending only on position. To rigorously confirm the correctness of such mathematical deduction, we seek the appropriate $\tilde{F}^{\nu\alpha}(x)$ by  the identification reasoning. To do it, we exploit (\ref{equation-37}) in order to calculate the two last terms appearing on the right side of (\ref{equation-36}). First, using the last result, it follows that Eqs. (\ref{equation-6}) and (\ref{equation-14}) give

\begin{eqnarray}\label{equation-38}
\big< \Gamma^{\nu\alpha\beta}(x) \left[x^{\mu}, \dot{x}_{\alpha}\right]\dot{x}_{\beta} \big> &=& \frac{i\hbar}{mR}\left\{\left(\eta^{\mu 0}-\frac{x^{\mu}}{R}\right)\Big<\left(1-\frac{x^{0}}{R}\right)^{2}\dot{x}^{\nu}\Big>\right.\nonumber\\ 
&+& \left.\Big<\left(1-\frac{x^{0}}{R}\right)^{2}\dot{x}^{\nu}\Big>\left(\eta^{\mu 0}-\frac{x^{\mu}}{R}\right)\right\},
\end{eqnarray}

\noindent which, by using  (\ref{equation-24}) where

\begin{equation}\label{equation-40}
\dot{x}^{\nu}\left(\eta^{\mu 0}-\frac{x^{\mu}}{R}\right)=\left(\eta^{\mu 0}-\frac{x^{\mu}}{R}\right)\dot{x}^{\nu}-\frac{1}{R}\left[\dot{x}^{\nu}, x^{\mu}\right],
\end{equation}

\noindent  reduces to

\begin{eqnarray}\label{equation-41}
\big< \Gamma^{\nu\alpha\beta}(x) \left[x^{\mu}, \dot{x}_{\alpha}\right]\dot{x}_{\beta} \big> &=&-\frac{2}{R}\left(1-\frac{x^{0}}{R}\right)^{-1}\left[x^{\mu}, \dot{x}^{0}\right]\dot{x}^{\nu} + \frac{i\hbar}{mR^{2}}\left(1-\frac{x^{0}}{R}\right)^{2}\left[x^{\mu}, \dot{x}^{\nu}\right]\nonumber\\
&&+\frac{2\hbar^{2}}{m^{2}R^{2}}\left(1-\frac{x^{0}}{R}\right)^{4}\left\{\eta^{\mu 0}\eta^{0\nu}-\frac{1}{R}\big(\eta^{0\nu}x^{\mu}+\eta^{\mu 0}x^{\nu}\big) + \frac{x^{\mu}x^{\nu}}{R^{2}}\right\}.
\end{eqnarray}

\noindent Next, taking into account (\ref{equation-6}) and (\ref{equation-37}), we get

\begin{eqnarray}\label{equation-42}
\big< \Gamma^{\nu\alpha\beta}(x)\left[x^{\mu}, \dot{x}_{\beta}\right] \dot{x}_{\alpha}\big>&=& \frac{2i\hbar}{mR}\eta^{\mu\nu}\Big<\left(1-\frac{x^{0}}{R}\right)\dot{x}^{0}\Big> - \frac{2i\hbar}{mR^{2}}\left\{\eta^{\mu 0}\Big<\left(1-\frac{x^{0}}{R}\right)x^{\nu}\dot{x}^{0}\Big> \right.\nonumber\\
&+& \left.\eta^{0 \nu}\Big<\left(1-\frac{x^{0}}{R}\right)x^{\mu}\dot{x}^{0}\Big>\right\}+\frac{2i\hbar}{mR^{3}}\Big<\left(1-\frac{x^{0}}{R}\right)x^{\mu}x^{\nu}\dot{x}^{0}\Big>
\end{eqnarray} 

\noindent which, after using (\ref{equation-12}), (\ref{equation-18}), (\ref{equation-19}) and (\ref{equation-23}), yields

\begin{eqnarray}\label{equation-43}
\big< \Gamma^{\nu\alpha\beta}(x)\left[x^{\mu}, \dot{x}_{\beta}\right] \dot{x}_{\alpha}\big>&=&-\frac{2}{R}\left(1-\frac{x^{0}}{R}\right)^{-1}\left[x^{\mu}, \dot{x}^{\nu}\right]\dot{x}^{0} + \frac{\hbar^{2}}{m^{2}R^{2}} \left(1-\frac{x^{0}}{R}\right)^{4} \left\{\left(\eta^{\mu\nu} + 2\eta^{\mu 0}\eta^{0\nu}\right) \right.\nonumber\\
&&\left. -\frac{3}{R}\big(\eta^{0\nu}x^{\mu}+\eta^{\mu 0}x^{\nu}\big) + \frac{3}{R^{2}}x^{\mu}x^{\nu}\right\}.
\end{eqnarray}

\noindent By summing (\ref{equation-41}) with (\ref{equation-43}) and identifying the final result with (\ref{equation-34}), we find

\begin{eqnarray}\label{equation-44}
\left[x^{\mu}, F^{\nu}\right]&=&\frac{i\hbar q}{m}F^{\mu\nu}(x)+ m\big< \Gamma^{\nu\alpha\beta}(x) \left( \left[x^{\mu}, \dot{x}_{\alpha}\right]\dot{x}_{\beta} + \dot{x}_{\alpha}\left[x^{\mu}, \dot{x}_{\beta}\right] \right)\big>,
\end{eqnarray}

\noindent which, by comparing it with (\ref{equation-36}), gives

\begin{equation}\label{equation-45}
\tilde{F}^{\nu\alpha}(x)[x^{\mu}, \dot{x}_{\alpha}]=\frac{i\hbar}{m}F^{\mu\nu}(x).
\end{equation}

\noindent  To go further, we put $\mu=0$ in the last relation in order to determine the final expression of $\tilde{F}^{\nu\alpha}(x)$. Indeed, using Eq. (\ref{equation-6}), it results

\begin{equation}\label{equation-46}
\frac{1}{R}\tilde{F}^{\nu}_{\alpha}x^{\alpha}=-\left(1-\frac{x^{0}}{R}\right)^{-3}F^{\nu 0} + \tilde{F}^{\nu 0}.
\end{equation}

\noindent which, by substituting it in (\ref{equation-45}), yields

\begin{equation}\label{equation-47}
\tilde{F}^{\nu\mu}=\left(1-\frac{x^{0}}{R}\right)^{-2}F^{\nu\mu} -\left(1-\frac{x^{0}}{R}\right)^{-3}\left(\eta^{\mu 0}-\frac{x^{\mu}}{R}\right)F^{\nu 0} + \eta^{\mu 0}\tilde{F}^{\nu 0}.
\end{equation}

\noindent By putting above $\nu=0$, it follows

\begin{equation}\label{equation-48}
\tilde{F}^{0\mu}=\left(1-\frac{x^{0}}{R}\right)^{-2}F^{0\mu}
\end{equation}

\noindent and then (\ref{equation-47}) takes the following final form

\begin{equation}\label{equation-49}
\tilde{F}^{\nu\mu}=\left(1-\frac{x^{0}}{R}\right)^{-2}F^{\nu\mu} + \frac{1}{R}\left(1-\frac{x^{0}}{R}\right)^{-3}\left(x^{\mu}- \eta^{\mu 0}x^{0}\right)F^{\nu 0}.
\end{equation}

\noindent To zeroth order in $1/R$, we get $\tilde{F}^{\nu\mu}=F^{\nu\mu}$. Finally, the substitution of (\ref{equation-37}) and (\ref{equation-49}) into (\ref{equation-35}) gives the following exact form of Lorentz force in Fick's nonlinear relativity

\begin{eqnarray}\label{equation-50}
F^{\nu}&=&q\Big<\left\{\left(1-\frac{x^{0}}{R}\right)^{-2}F^{\nu\mu} + \frac{1}{R}\left(1-\frac{x^{0}}{R}\right)^{-3}\left(x^{\mu}- \eta^{\mu 0}x^{0}\right)F^{\nu 0}\right\}\dot{x}_{\mu}\Big>\nonumber\\ 
&-&\frac{2m}{R}\Big< \left(1-\frac{x^{0}}{R}\right)^{-1} \dot{x}^{0} \dot{x}^{\nu}\Big> + G^{\nu}(x).
\end{eqnarray}

\noindent Analysing the above result, we remark that the particle mass acts so that for any two particles with the same charge, submitted to the same electromagnetic field, will not necessarily feel the same force if their masses are different. Surprisingly, it is notable that mass and charge emerge independently of each other in two distinct terms  where the new effect is completely independent of  electromagnetic field. As one consequence, even for a neutral particle, moving without the influence of electromagnetic field, a particular kind of force acts on it (gravitational-type Lorentz force).  Using Eqs. (\ref{equation-46}) and (\ref{equation-48}), we can check that Eq. (\ref{equation-50}) reproduces the first order approximation previously found in \cite{Takka-Bouda-Foughali-1}. In the temporal limit where $x^{0}=R$, it is obvious that no force is generated $(F^{0}=0,\hspace*{0.1cm}\vec{F}=\vec{0})$.

 
 

\section{Conclusion}

This work constitutes another effort to better understand the impact of the variation of the fundamental physical constants on the basic laws of electrodynamics. Thus, to find the first-order approximation of Lorentz force in the context of Fock’s nonlinear relativity, we have developed in Ref. \cite{Takka-Bouda-Foughali-1} an adequate procedure for such derivation.
Building on this latter, we have highlighted in Ref. \cite{Takka-Bouda-2} that an iterative method could be used in order to move towards the higher-order terms. However, it is turned out that this reasoning quickly becomes obsolete as we seek to increase the precision (even for the second order, the calculation is already too complicated). To overcome this difficulty, we have succeeded here not only to simplify the first procedure but we also went further by proposing an elegant exact study. In doing so, we have shown that the aforementioned force is decomposed into two different contributions where each effect is reinforced in its proper nature by a separability of variables. Indeed, in comparison with the generalized Lorentz force found in Ref. \cite{Harikumar-Juric-Meljanac-1}, valid up to first order in the deformation parameter $\kappa$, we emphasize that mass and charge intervene linearly and independently of each other in our work but as a product in the $\kappa$-deformed one. Interpreting the two effects appearing in the final result, the first term linear in velocity, depending on charge but not on mass, is named the electric-type Lorentz force, the second one proportional to the square of the velocity, depending
on mass but not on charge, is interpreted as the gravitational-type Lorentz force. Consequently, more homogeneity regarding the cause and effect relationship takes place as a new important particularity in our context. More specifically, it thus seems that the similarity between electromagnetism and gravity is more visible in FNLR than in $\kappa$-Minkowski space–time.

To summarize a little bit better what has been shown until now, it is very useful to shed light on some important points. First of all, the reasoning developed above is nothing else than a logical continuation of the two last papers where the theoretical background is constructed in a coherent way. Indeed, our approach is based on our version of Feynman’s proof, itself based on both the relativistic version of this latter and the use of the momentum valid in the absence of the electromagnetic field. Mathematically, all above calculations can be viewed as a symmetrization mechanism allowing an identification work. As one perspective, the ordinary Lagrangian formulation could be envisaged as an approximate study since the calculations seem to be complicated even for the first order, e.g. DSR \cite{Pramanik-Ghosh}. Generally speaking, the standardization of our resolution procedure is technically possible but only conditioned by the mathematical development taking into account the specificities of each new context. Secondly, in addition to the exact nature of our predictions, some new particularities concerning both the content and form can be illustrated. For example, the separability between mass and charge in the R-deformed Lorentz force is in accordance with the independence of the two physical characteristics in standard physics. From a practical point of view, the relevance of our studies is the proposition of new theoretical challenges to check the validity of the invariance of the speed of light, the homogeneity of space and the uniformity of time in the cosmological context \cite{Bouda-Foughali-1}. In an equivalent way, the theoretical results highlighted in FNLR, \cite{Takka-Bouda-Foughali-1}, \cite{Takka-Bouda-2}, \cite{Foughali-Bouda-1},\cite{Foughali-Bouda-2} provide a new series of occasions to test one of the most fundamental concepts of physics, namely, the conservation of the energy at the scale of the universe \cite{Bouda-Foughali-1}. To examine the scientific soundness of our predictions, it is obvious that the empirical or observational studies must be focused on the
physical phenomena of the large scale in space and/or time. In the same order of ideas, we emphasize that all our final results take a very condensed form despite the fact that the calculations are very long. Also, the non-use of any other assumption beside the relativistic Newton’s law, making possible the reproduction of the usual Lorentz force, constitutes in itself another technical and mathematical prowess. Before finishing, let us take the most simplest case characterized by the classical
static limit. In fact, by considering two identical charged particles $q$ at rest and separated by a distance $r$, it follows that $\dot{x}^{0}=1$ and $\dot{x}^{i}=0$ which implies that Eq. (\ref{equation-50}) reduces to

\begin{equation}\label{equation-51}
\vec{F}=q\left(1-\frac{x^{0}}{R}\right)^{-2}\vec{E}.
\end{equation}

In contrast to $\kappa$-Minkowski space–time, where the first extended form of Coulomb’s law depends on mass and charge, \cite{Harikumar-Juric-Meljanac-1} we note here an exclusive dependence on charge in our exact result. This suggests that the gravitational-type effect is mainly related to the movement which supports even more the conclusion drawn just before. Based
on these consequences, many other analogies and more detailed analysis could be envisaged in other works.

 
 

\end{document}